\documentclass[fleqn,10pt]{wlscirep}
\usepackage[utf8]{inputenc}
\usepackage[T1]{fontenc}
\usepackage{amsmath, amssymb}
\usepackage{xcolor,comment}
\usepackage{geometry}
\usepackage{graphicx}
\usepackage{todonotes}
\usepackage{amsthm}
\usepackage{algorithm}
\usepackage{algpseudocode}
\usepackage{hyperref}
\hypersetup{
     colorlinks= true,
     linkcolor = blue,
     filecolor = blue,
     citecolor = blue,      
     urlcolor  = blue,
     }
\usepackage{cleveref}

\title{Community Detection in Hypergraphs via Mutual Information Maximization}

\author[1]{J\"{u}rgen Kritschgau}
\author[6]{Daniel Kaiser}
\author[2,+]{Oliver Alvarado Rodriguez}
\author[3,+]{Ilya Amburg}
\author[4,+]{Jessalyn Bolkema}
\author[5,+]{Thomas Grubb}
\author[7,+]{Fangfei Lan}
\author[8,+]{Sepideh Maleki}
\author[9]{Phil Chodrow}
\author[3,*]{Bill Kay}

\affil[1]{Carnegie Mellon University, Department of Mathematical Sciences, Pittsburgh, PA 15213, USA}
\affil[2]{New Jersey Institute of Technology, Department of Computer Science, Newark, NJ 07102, USA}
\affil[3]{Pacific Northwest National Laboratory, Richland, WA 99354, USA }
\affil[4]{California State University, Dominguez Hills, Department of Mathematics, Carson, CA 90747, USA}
\affil[5]{University of California San Diego, San Diego, CA 92093, USA}
\affil[6]{Indiana University, Department of Informatics, Bloomington, IN 47408, USA}
\affil[7]{University of Utah, Scientific Computing and Imaging Institute, Salt Lake City, UT 84112, USA}
\affil[8]{University of Texas at Austin, Department of Computer Science, Austin, TX 78712, USA}
\affil[9]{Middlebury College, Department of Computer Science, Middlebury, VT 05753, USA}
\affil[*]{Correspondence to: william.kay@pnnl.gov}

\affil[+]{Equal contributions}

\date{July 2023}

\newcommand{\Hy}{\mathsf{H}}

\newcommand{\vc}{\textbf{c}}

\newcommand{\abs}[1]{\left|#1\right|}

\newcommand{\cX}{\mathcal{X}}
\newcommand{\cY}{\mathcal{Y}}
\newcommand{\Z}{\mathbb{Z}}

\newcommand{\braces}[1]{\left\{#1\right\}}

\DeclareMathOperator*{\argmax}{arg\,max}
\DeclareMathOperator*{\argmin}{arg\,min}

\definecolor{philcolor}{rgb}{0.9,0.7,1}

\definecolor{jesscolor}{rgb}{0.67,0.88,0.69}

\definecolor{billcolor}{rgb}{1,.88,0.21}

\definecolor{juergencolor}{rgb}{0.67,0.9,0.93}

\definecolor{danncolor}{rgb}{0.85,0.2,0.6}

\definecolor{ilyacolor}{rgb}{0.2,0.85,0.6}

\theoremstyle{definition}
\newtheorem{dfn}{Definition}[section]

\begin{abstract}
    
    The hypergraph community detection problem seeks to identify groups of related nodes in hypergraph data. 
    We propose an information-theoretic hypergraph community detection algorithm which compresses the observed data in terms of community labels and community-edge intersections. 
    This algorithm can also be viewed as maximum-likelihood inference in a degree-corrected microcanonical stochastic blockmodel. 
    We perform the inference/compression step via simulated annealing. 
    Unlike several recent algorithms based on canonical models, our microcanonical algorithm does not require inference of statistical parameters such as node degrees or pairwise group connection rates. 
    Through synthetic experiments, we find that our algorithm succeeds down to recently-conjectured thresholds for sparse random hypergraphs. 
    We also find competitive performance in cluster recovery tasks on several hypergraph data sets. 
    
\end{abstract}
\begin{document}

\maketitle

\section{Introduction}

% community detection in graphs and hypergraphs, many methods
% information theoretic framing, Rosvall-Bergstrom
% our contribution

The network clustering task asks us to identify sets (``clusters'') of vertices in a network with the property that vertices in each cluster are related to each other in some way that they are not related to vertices in other clusters. 
In various disciplines, the graph clustering task may also be called \emph{network partitioning} or \emph{community detection}. 
A large number of methods have been developed for clustering dyadic networks, in which relationships exist between pairs of vertices. 
Such dyadic networks can be represented as graphs. 
Techniques for graph clustering include spectral methods, greedy optimization methods, and methods based on statistical inference, with many theoretical connections across these categories.\cite{newman2018networks} 

Much recent work has emphasized the importance of polyadic interactions---interactions between groups of two or more entities---in complex systems.\cite{bickWhatAreHigherorder2021,torresWhyHowWhen2021} 
Such interactions can often be modeled as edges in a hypergraph. 
Hypergraphs pose both opportunities and challenges for clustering algorithms. 
On the one hand, the richer representation of relationships offered by hypergraphs can in some cases produce superior performance when compared to graph methods applied to the same data. 
On the other hand, the flexibility implied by arbitrary edge sizes can lead both computational and statistical pitfalls. 
There are many extant approaches to hypergraph clustering including spectral methods,\cite{keCommunityDetectionHypergraph2019} methods based on combinatorial optimization,\cite{chodrow2021generative,veldtHypergraphCutsGeneral2022,schlagHighQualityHypergraphPartitioning2022} and methods based on statistical inference in both single-membership and mixed-membership generative models.\cite{chodrowNonbacktrackingSpectralClustering2022,ruggeriGeneralizedInferenceMesoscale2023} 

In this paper, we offer a hypergraph clustering algorithm with information-theoretic foundations. 
This algorithm extends a method proposed by Rosvall and Bergstrom  for graph clustering.\cite{rosvallInformationtheoreticFrameworkResolving2007} 
Their approach begins by regarding a proposed clustering of a graph as a lossy \emph{compression} of the graph. 
The aim is then to form a compression that, for a fixed storage size, is maximally informative of the original graph structure. 
They formulate this criterion in terms of maximization of mutual information, or, equivalently, minimization of a certain entropy functional. 
They then use simulated annealing to perform the minimization. 
This approach is equivalent to maximum-likelihood estimation in a microcanonical graph stochastic blockmodel,\cite{peixotoNonparametricBayesianInference2017} and may thus also be viewed as a statistical inference method. 

Our proposed method extends the algorithm of Rosvall-Bergstrom  algorithm by (a) formulating the entropy functional on the more combinatorially complex set of hypergraphs and (b) incorporating a ``degree-correction'' \cite{karrerStochasticBlockmodelsCommunity2011,peixotoNonparametricBayesianInference2017} to account for heterogeneity of node degrees. 
\Cref{sec:methods} contains a description of the entropy functional, its information theoretic foundations, and the simulated annealing algorithm we use to locally minimize the entropy.
In \Cref{sec:synthetic}, we demonstrate our algorithm on several synthetic data sets, finding experimental suggestion that the algorithm succeeds down to the sparse detectability limit conjectured by Chodrow et al. (2023).\cite{chodrowNonbacktrackingSpectralClustering2022} 
In \Cref{sec:empirical}, we conduct experiments on several empirical data sets, finding performance competitive with extant graph and hypergraph methods. 
We close in \Cref{sec:discussion} with discussion of our findings and suggestions for future work.

\section{Methods} \label{sec:methods}

We treat the hypergraph clustering problem as an information-theoretic compression problem, in which the aim is to find a maximally informative, clustered description of the hypergraph structure. 
In this section, we introduce the core technical ideas needed to describe this approach: hypergraph compressions, information, and entropy. 

\subsection{Hypergraph Compression}

Let $\Hy$ be a hypergraph with edge set $E=E(\Hy)$ and vertex set $V=V(\Hy)$.
Suppose $\braces{C_i}_{i=1}^m$ is a partition of $V$ into $m$ clusters. 
If $\lambda = (\lambda_1, \ldots, \lambda_m)\in \mathbb N^{m}$, we say an \emph{edge $A\in E$ is of $\lambda$-type} if $\abs{A\cap C_i}=\lambda_i$ for $1\leq i \leq m$. 
That is, $\lambda_i$ counts the number of nodes in edge $A$ in cluster $C_i$. 
We denote by $E_\lambda$ the set of all edges of $\lambda$-type. 

\begin{dfn}[Hypergraph Compression]
    A \emph{compression of $\Hy$ into $m$ clusters} is a pair $\gamma = (\braces{C_i}_{i=1}^m,\braces{e_\lambda}_{\lambda \in \mathbb N^{m}})$ such that 
    \begin{itemize}
        \item $\braces{C_i}_{i=1}^m$ is a partition of $V$, and 
        \item $\braces{e_\lambda}_{\lambda \in \mathbb N^{m}}$ is a collection indexed by $\lambda$, where $e_\lambda$ is the number of $\lambda$-type edges in $\Hy$.   
    \end{itemize}
    We say that $\Hy$ and $\gamma$ are \textit{compatible} if $\gamma$ is a compression of $\Hy$. 
    We let $\mathbb{H}(\gamma)$ be the set of all hypergraphs compatible with a fixed $\gamma$, and let $Z(\gamma) = \abs{\mathbb{H}(\gamma)}$. 
    We also let $\Gamma(\Hy)$ denote the set of compressions compatible with $\Hy$. 
\end{dfn}

The collection of clusters $\braces{C_i}_{i=1}^m$ may be equivalently represented as an assignment vector, as e.g. used by Rosvall and Bergstrom.\cite{rosvallInformationtheoreticFrameworkResolving2007} 
When convenient, we may refer to an assignment vector $\vc\in \braces{1,\dots,m}^{V}$ where $c_v=i$ if and only if $v\in C_i$. 
Similarly, if $\Hy$ is a simple graph, then $\braces{e_\lambda}_{\lambda \in \mathbb N^{m}}$ reduces to the module matrix in Rosvall and Bergstrom's formulation. \cite{rosvallInformationtheoreticFrameworkResolving2007} 

In applications, it is useful to also incorporate the node degree sequence into the compressed representation of the hypergraph. 
Let $\braces{d_v}_{v\in V}$ be the degree sequence of nodes in $\Hy$. 
\begin{dfn}[Degree-Corrected Hypergraph Compression]
    A \emph{compression of $\Hy$ into $m$ clusters with degrees} is a triple $\gamma = (\braces{C_i}_{i=1}^m,\braces{e_\lambda}_{\lambda \in \mathbb N^{m}},\braces{d_v}_{v\in V})$. 
\end{dfn}
Explicitly incorporating the degree sequence into the compression is the analogue of degree-correction in canonical stochastic blockmodels. \cite{karrerStochasticBlockmodelsCommunity2011} 
We discuss the connection to stochastic blockmodels in \Cref{sec:max-likelihood-connection}. 
Throughout the remainder of this paper, we let $\Gamma(\Hy)$ denote the set of all compressions of a fixed hypergraph $\Hy$, describing in context when necessary whether the space of compressions includes degrees.

\begin{figure}[H]
\centering
    \includegraphics[width=0.5\linewidth]{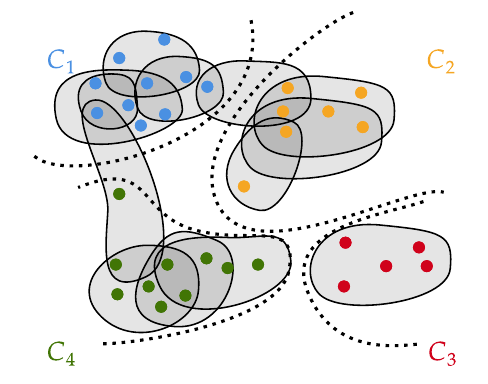}
    \caption{Example of a hypergraph whose vertices have been partitioned into four clusters.}
    \label{fig:cluster_example}
\end{figure}

\subsection{Information and Entropy}

For a given hypergraph, our aim is to select a ``maximally informative'' compression. 
We define the information content of a compression in terms of Shannon entropy.\cite{shannon1948mathematical} 
Our definitions follow the formulation of Cover and Thomas (2012).\cite{cover2012elements} 
Let $X$ and $Y$ be discrete random variables with joint distribution $p(x, y)$ over an alphabet $\cX \times \cY$. 
\begin{dfn}[Marginal, Joint, and Conditional Entropies]
    The \emph{marginal entropy} (or simply \emph{entropy}) of the random variable $X$ is 
    \begin{align*}
        H(X) \triangleq - \sum_{x \in \cX}p(x) \log p(x)\;. 
    \end{align*}
    The \emph{joint entropy} of $X$ and $Y$ is 
    \begin{align*}
        H(X, Y) \triangleq - \sum_{x \in \cX}\sum_{y \in \cY} p(x, y) \log p(x, y)\;. 
    \end{align*}
    The \emph{conditional entropy} of $Y$ given $X$ is 
    \begin{align*}
        H(Y|X) = - \sum_{x \in \cX}\sum_{y \in \cY} p(x, y) \log p(y|x)\;.
    \end{align*}
\end{dfn}
The entropy $H(X)$ can be viewed as a measure of spread for the discrete random variable $X$. 
It is maximized with respect to the distribution $p$ by the uniform distribution $p(x) = \frac{1}{\abs{\cX}}$, in which case $H(X) = \log \abs{\cX}$. 
The joint entropy $H(X, Y)$ is similarly a measure of spread for the joint distribution $p(X, Y)$. 
The conditional entropy $H(Y|X)$ is the expected spread in the distribution $p(y|x)$ across realizations of $x$, as highlighted by the formula 
\begin{align*}
    H(Y|X) &= - \sum_{x \in \cX}\sum_{y \in \cY} p(x)p(y|x) \log p(y|x) \\ 
    &= -\sum_{x \in \cX}p(x)H(Y|X = x)\;.
\end{align*}

\begin{dfn}[Mutual Information] \label{def:mutual-info}
    The {\em mutual information} of $X$ and $Y$ is given by:
    \begin{align*}
        I(X;Y) &\triangleq H(X) - H(X|Y)\\
        &=H(Y) - H(Y|X)\;.
    \end{align*}
\end{dfn}
Other definitions of the mutual information exist under which \Cref{def:mutual-info} is a theorem rather than a definition. 
Treating $H(X)$ as a measure of uncertainty about $X$, and $H(X|Y)$ as a measure of uncertainty about $X$ conditional on knowing the value of $Y$, the mutual information measures how much knowledge of $Y$ \emph{reduces} uncertainty in $X$.

\subsection{Information Maximization as Counting}
% Direct analog to RB is BLAH (no repeated edges) and on synthetic data where every edge has the same degree in expectation, this performs well. As RB point out, their approach does not do well when there is high variance among degrees, and similarly, the above count does not handle degree variation well. 

Our aim is to choose a compression $\gamma$ that is maximally informative about the structure of the hypergraph $\Hy$. 
%To formalize this, we model $\Hy$ as being drawn from a distribution 
Let $\Gamma$ be a set of possible compressions and, for each $\gamma \in \Gamma$, $p(\cdot ~|~\gamma)$ be uniform on $\mathbb{H}(\gamma)$. 
In practice, we usually take $\Gamma = \Gamma(\Hy_0)$ to be the set of all compressions compatible with an observed hypergraph $\Hy_0$.  
We assume an unspecified prior $q$ over $\Gamma$ which we will soon optimize. We model $\Hy$ as being drawn from a distribution: 

\begin{align*}
    p(\Hy) = \sum_{\gamma \in \Gamma} p(\Hy ~|~ \gamma)q(\gamma)\;.
\end{align*}

We form the compression $\gamma$ and sample a new hypergraph $\Hy'$ from the distribution $p(\cdot~|~\gamma)$. 
We can think of this process as describing the hypergraph $\Hy$ by transmitting the compression $\gamma$ to a stranger who does not observe $\Hy$ itself. 
The stranger then forms a guess $\Hy'$ about the structure of the hypergraph described by the compression. 

We seek a distribution $q$ over $\Gamma$ that maximizes the mutual information between $\Hy$ and $\Hy'$: 
\begin{align} \label{eq:mutual-info-max}
    q = \argmax_q  I(\Hy; \Hy') \quad \text{such that} \quad \Hy' \sim p(\cdot ~|~\gamma) \quad \text{and} \quad \gamma \sim q\;. 
\end{align}
To simplify this problem, we first observe that, by construction,  $\Hy$ and $\Hy'$ are independent conditioned on $\gamma$:
\begin{align*}
    p(\Hy, \Hy'~|~\gamma) = p(\Hy'~|~\Hy, \gamma)p(\Hy~|~\gamma) = p(\Hy'~|~ \gamma)p(\Hy~|~\gamma)\;.
\end{align*} 
The last equality reflects the fact that, once $\gamma$ is transmitted, the signal receiver does not have any other access to $\Hy$ when generating the guess $\Hy'$. 
Now applying the chain rule of mutual information, we have 
\begin{align*}
    I(\Hy;\Hy') = I(\Hy; \gamma, \Hy') - I(\Hy; \Hy'|\gamma)\;.
\end{align*}
By conditional independence, $I(\Hy; \Hy'|\gamma) = 0$ and $I(\Hy; \gamma, \Hy') = I(\Hy; \gamma)$. 
It follows that 
\begin{align*}
    I(\Hy; \Hy') = I(\Hy; \gamma) = H(\Hy) - H(\Hy~|~\gamma)\;. 
\end{align*}
Since the first term does not depend on $\gamma$, we can ignore it in the optimization over $q$, and our reduced problem becomes 
\begin{align*}
    q = \argmin_q H(\Hy~|~\gamma) \quad \text{such that} \quad \gamma \sim q\;.
\end{align*}
Expanding the conditional entropy yields 
\begin{align*}
    H(\Hy ~|~ \gamma) &= \sum_{\gamma \in \Gamma}\sum_{\Hy \in \mathbb{H}(\gamma)} p(\Hy, \gamma) \log p(\Hy ~|~\gamma) \\ 
                      &= \sum_{\gamma\in \Gamma} q(\gamma) \sum_{\Hy \in \mathbb{H}(\gamma)} p(\Hy ~|~ \gamma) \log p(\Hy ~|~\gamma)\;.
\end{align*} 
This expression makes clear that the optimal $q$ concentrates all its mass on values $\gamma$ that minimize the entropy of the distribution $p(\cdot ~|~ \gamma)$. 
But since $p(\cdot ~|~ \gamma)$ is uniform, the entropy of this distribution is simply $\log Z(\gamma)$, where $Z(\gamma) = \abs{\mathbb{H}(\gamma)}$ is the number of hypergraphs compatible with $\gamma$.   
Thus, after observing a data hypergraph $\Hy_0$ and setting $\Gamma = \Gamma(\Hy)$, our original mutual information maximization problem \cref{eq:mutual-info-max} reduces to the problem 
\begin{align}
    \hat{\gamma} = \argmin_{\gamma \in \Gamma(\Hy_0)} Z(\gamma)\;. \label{eq:min-ent}
\end{align}
That is, the maximally informative compression $\gamma$ of a given hypergraph $\Hy_0$ is the compression that is compatible with $\Hy_0$ and minimizes the size of $\mathbb{H}(\gamma)$. 
We can think of $\gamma$ as a description of $\Hy_0$ that minimizes the number of alternative hypergraphs $Z(\gamma)$ which could also be described by $\gamma$. 

\subsection{Relation to Maximum-Likelihood Estimation} \label{sec:max-likelihood-connection}

    The entropy minimization problem \cref{eq:min-ent} and maximum-likelihood estimation arise from particular stochastic blockmodel. 
    Recall the conditional data generating distribution $p(\cdot ~|~\gamma)$, which is uniform over the set $\mathbb{H}(\gamma)$ of all hypergraphs compatible with the compression $\gamma$: 
    \begin{align*}
        p(\Hy ~|~\gamma) = \begin{cases}
            \frac{1}{Z(\gamma)} \quad \Hy \in \mathbb{H}(\gamma) \\ 
            0 \quad \text{otherwise.}
        \end{cases}
    \end{align*}
    We can then equivalently write our minimum-entropy problem as 
    \begin{align}
        \gamma = \argmin_{\gamma \in \Gamma(\Hy_0)} Z(\gamma) = \argmax_{\gamma \in \Gamma(\Hy_0)} \frac{1}{Z(\gamma)} = \argmax_{\gamma \in \Gamma(\Hy_0)} p(\Hy~|~\gamma)\;.  \label{eq:max-likelihood}
    \end{align}
    Since $\gamma$ itself contains cluster memberships and edge-cluster intersections, $p(\cdot~|~\gamma)$ can be viewed as a microcanonical hypergraph stochastic blockmodel,  generalizing known microcanonical models for graphs.\cite{peixotoNonparametricBayesianInference2017} 
    The mutual information maximization \cref{eq:mutual-info-max}, the entropy minimization \cref{eq:min-ent}, and the maximum-likelihood problem \cref{eq:max-likelihood} are all equivalent ways to describe our inference problem. 

Rosvall and Bergstrom \cite{rosvallInformationtheoreticFrameworkResolving2007}  count the number of graphs $G$ that admit $\gamma = (\braces{C_i}_{i=1}^m, \mathbb M)$ as a compression, where $\braces{C_i}_{i=1}^m$ is a partition of the vertex set of $G$ and each entry of the module matrix $\mathbb M_{i,j}$ enumerates the number of edges between cluster $i$ and $j$, as follows:  
\begin{equation*}
Z(\gamma) =  \prod_{i<j} \binom{\abs{C_i}\abs{C_j}}{\mathbb M_{i,j}}\prod_{i=1}^m \binom{\binom{\abs{C_i}}{2}}{\mathbb M_{i,i}}. \tag{Equation 4 of Rosvall and Bergstrom \cite{rosvallInformationtheoreticFrameworkResolving2007}}
\end{equation*}
Our aim is to maximize the mutual information between a hypergraph $\Hy$ and its compression. 
To do this via \Cref{eq:min-ent}, we need to evaluate $Z(\gamma)$, the number of hypergraphs compatible with the compression $\gamma$. 
If we restrict to \emph{simple} hypergraphs, which do not have multiple edges, then 
\begin{align*}
Z(\gamma) =  \prod_{\lambda\in \mathbb N^m}\binom{\prod_{i=1}^m \binom{\abs{C_i}}{\lambda_i}}{e_\lambda}. 
\end{align*}
We remark that the (a priori) infinite limit exists, as all but finitely many $\lambda$ are $\mathbf{0}$. 
Here, the expression $\prod_{i=1}^m \binom{\abs{C_i}}{\lambda_i}$ counts the number of ways to choose the appropriate number of vertices from each of the $m$ clusters for inclusion in one $\lambda$-edge, from which we select $e_\lambda$ edges without repetition to realize. 

If we instead consider multi-hypergraphs, in which multiple edges are permitted, then there are \[\left(\prod_{i=1}^m \binom{\abs{C_i}}{\lambda_i}\right)^{e_\lambda} =\prod_{A \in E_\lambda}\prod_{i=1}^m \binom{\abs{C_i}}{\lambda_i} \] ways to select the $e_\lambda$ edges from among all possible edges of type $\lambda$. 
It follows in this case that 
\begin{align*} 
Z(\gamma)=  \prod_{\lambda\in \mathbb N^m} \prod_{A\in E_\lambda} \prod_{i=1}^m  \binom{\abs{C_i}}{\lambda_i}.
\end{align*}
Noting that,  $\lambda_i=\abs{A \cap C_i}$ if $A \in E_\lambda$, we can rewrite this expression as  
\begin{align*}
Z(\gamma)  =   \prod_{\lambda\in \mathbb N^m} \prod_{A\in E_\lambda} \prod_{i=1}^m \binom{\abs{C_i}}{\abs{A\cap C_i}}  = \prod_{A\in E} \prod_{i=1}^m \binom{\abs{C_i}}{\abs{A\cap C_i}}.
\end{align*}
Notably, this final expression is not organized according to edge type. 

% Enter: degree correction. Based on the DC-SBM of Tiago, 

% - Soft clustering? Ranked Choice Clustering? (if we have results)

% -Degree correction. Highlight this.

\subsection{Degree Adjusted Entropy} \label{sec.deg.adjustement}
In this section we vary the compression to allow for specification of a degree sequence in the hypergraph. In doing so, we will obtain a new entropy based objective function to minimize. As in the previous section, this entropy will be inspired by a hypergraph counting task. 

We now consider degree-corrected compressions of the form $\gamma = (\braces{C_i}_{i=1}^m,\braces{e_\lambda}_{\lambda \in \mathbb N^{m}},\braces{d_i}_{i\in V})$. 
We again let $Z(\gamma)$ denote the number of hypergraphs compatible with $\gamma$ as a degree-corrected compression.
We again seek to maximize mutual information by minimizing $Z(\gamma)$, which again requires a formula for $Z(\gamma)$. 

Let \[e_i = \sum_{\lambda \in \mathbb N^m} \lambda_ie_\lambda \] for $1\leq i \leq m$ denote the degree sum of vertices in cluster $C_i$. 
In what follows, we treat degrees as distinguishable ``stubs" hanging off of vertices. 
We imagine constructing a hypergraph $\Hy$ with the desired compression $\gamma$ through the following process:
\begin{enumerate}
    \item First, assign the available stubs within each cluster $C_i$ to the $\lambda$-types to which they will contribute.
    \item Second, for each $\lambda$-type:
    \begin{enumerate}
        \item for each $1\leq i \leq m$, group the assigned stubs from cluster $C_i$ into packets of size $\lambda_i$, then 
        \item combine the packets into edges of $\lambda$-type. 
    \end{enumerate}
\end{enumerate}
To count the number of hypergraphs compatible with $\gamma$, it suffices to count the number of possible $\lambda$-type assignments, $a(\gamma)$, from which to choose in Step 1, and then for each $\lambda \in \mathbb N^m$, the number of possible packets, $p_\lambda(\gamma)$, from which to choose in Step 2(a) and the number possible combinations of these packets into edges, $c(\gamma)$, in Step 2(b). 

%If there are $a(\gamma)$ ways to assign stubs in step 1, $g_i(\gamma)$ ways to group the assigned stubs from cluster $C_i$ into packets in step 2(a), and $c(\gamma)$ ways to combine packets into edges in step 2(b), then the total number of constructed compatible hypergraphs is $Z(\gamma)= a(\gamma)c(\gamma)$, where $c(\gamma)$ is a function of the $g_i(\gamma)$'s. 
The first assignment step can be done in
\begin{equation}  
   a(\gamma)= \prod_{i=1}^m \binom{e_i}{\dots,\lambda_ie_\lambda,\dots}.\label{eq.count1}
\end{equation}
possible ways, where the lower portion of the multinomial coefficient ranges over all $\lambda\in \mathbb N^m$. 

%\pcinline{Can we revise so that these displayed expressions have verbs? E.g. equations etc? Could be accomplished by naming the things to be counted in each step.} %took a stab at it! - JGB

To proceed with the second step, suppose $\lambda$ is fixed. 
Notice that each edge of $\lambda$-type requires $\lambda_i$ degrees from cluster $i$. 
Furthermore, recall that in step 1. we allocated $\lambda_ie_\lambda$ degrees for the purpose of construction $\lambda$-type edges. 
We can group the $\lambda_ie_\lambda$ degrees into packets of size $\lambda_i$ in $\binom{\lambda_ie_\lambda}{\dots, \lambda_i,\dots}$ ways, where the lower portion of the multinomial coefficient is repeated $e_\lambda$ times. 
Note that the packets produced by multinomial coefficients are \textbf{ordered}, which we will account for later. 
Repeating this process for each cluster completes Step 2(a) and can be done in a total of \begin{equation} p_\lambda(\gamma)=\prod_{i=1}^m \binom{\lambda_ie_\lambda}{\dots, \lambda_i,\dots}\label{eq.count2a}\end{equation}
ways. 

There is a natural way to combine packets into edges; simply take the first packet from each cluster to produce the first edge, then take the second packet from each cluster to produce the second edge, and so on (Note that if $\lambda_i=0$ we pretend like there is an infinite stream of empty packets). 
Notice that the same set of edges can be produced in $e_\lambda!$ ways. 
We account for this by dividing our count by $e_\lambda!$ (and this resolves the fact that the multinomial coefficients produced ordered packets). 
This essentially finishes step 2(b), which when combined with the expression from (\ref{eq.count2a}) for each $\lambda$ produces 
\begin{equation}
c(\gamma)=      \prod_{\lambda\in \mathbb N^m}\left( (e_\lambda!)^{-1}(p_\lambda(\gamma)\right) = \prod_{\lambda\in \mathbb N^m}\left( (e_\lambda!)^{-1}\prod_{i=1}^m \binom{\lambda_ie_\lambda}{\dots, \lambda_i,\dots}\right).
    \label{eq.count2}
\end{equation}

Therefore, combining expression (\ref{eq.count1}) and (\ref{eq.count2}), and forgetting the degree stub labels gives
\begin{align}
    Z(\gamma)   & = a(\gamma) c(\gamma)\nonumber \\
                & = \prod_{i=1}^m \binom{e_i}{\dots,\lambda_ie_\lambda,\dots}\prod_{\lambda\in \mathbb N^m}\left( (e_\lambda!)^{-1}\prod_{i=1}^m \binom{\lambda_ie_\lambda}{\dots, \lambda_i,\dots}\right)%\left(\sum_{v\in V} d_v!\right)^{-1}
    \nonumber \\
    &=\frac{\prod_{i=1}^m e_i!}{\left(\prod_\lambda e_\lambda!\right)\left(\prod_{\lambda}\prod_{i=1}^m (\lambda_i!)^{e_\lambda}\right)}.
    %\left(\sum_{v\in V}d_v!\right)^{-1}. 
    \label{eq.degree.adjusted.entropy}
\end{align}

An important remark is that we have technically counted hypergraphs $\Hy$ where we allow vertices to appear multiple times in an edge.
By distinguishing the stubs attached to each vertex from each other, we have also overcounted hypergraphs with parallel hyperedges. 
The expression \Cref{eq.degree.adjusted.entropy} is therefore an \emph{approximation} of the exact degree-corrected entropy. 
The quality of this approximation depends on the statistical prevalence of multiple vertex inclusions and parallel hyperedges.\cite{chodrowConfigurationModelsRandom2020}
In graphs with fixed degree sequences, it is known that, provided that the low-order moments of the degree sequence remain constant as the number of nodes grows large (i.e. in the ``large, sparse limit''), the number of multiple inclusions and parallel edges is concentrate around constants that depend on moments of the degree sequence.\cite{angelLimitLawsSelfloops2019}
It follows that the proportion of edges with multiple node inclusions or with parallel approaches zero in the limit. 
We are unaware of formal proofs of similar results for hypergraphs, or for graphs with community structure. 
We conjecture that the same heuristic should roughly hold: provided that the degree sequence and edge-size sequence of the hypergraph have low-order moments that are sufficiently small relative to the number of nodes, the approximate entropy will be very close to the exact entropy.

% The number of hypergraphs with loops that we count constitute a lower order term, and will not affect our results asymptotically. 
% However, by allowing these looped edges, we simplify the count in step 2(a) considerably since we don't have to be careful about grouping degrees from the same vertex into a single packet. 

% However, to precisely count hypergraphs with specified degree sequence is a difficult combinatorial goal in the general setting. We will ease the count in two ways. First, we allow edges in the hypergraph to have vertices with multiplicity; this is equivalent to allowing loops in dyadic networks. Second, we will treat the degree of a vertex as being assigned to distinguishable \emph{stubs} originating from that vertex. In that way, the two networks below are treated as different for counting's sake.

In light of allowing vertices to appear multiple times in an edge and the form of Equation \ref{eq.degree.adjusted.entropy}, it is tempting to assume that the  degree sequence of $\Hy$ does not impact the entropy calculation. 
This is partially correct.
The degrees matter up to cluster assignment; which is to say that the entropy calculation cares about the total degrees of the clusters, but not how the degrees are distributed within the clusters. 
However, the particular degree sequence of $\Hy$ does influence how the entropy calculation acts across the whole state space. 
When comparing two cluster assignments that differ in only one vertex $v$,  the degree of vertex $v$ essentially accounts for the difference in entropy of the two cluster assignments.
In other words, the degree of $v$ determines how the total degrees of clusters change when we change the cluster assignment of $v$; This in turn, determines how entropy changes. 

%When implementing the simulated annealing process, one can forgo the normalization by $\small\left(\sum_{v\in V} d_v!\right)^{-1}.$
%Any compression we consider will treat $\braces{d_v}_{v\in V}$ as observed data and require the same normalization.
%Furthermore, the Metropolis-Hastings Algorithm determines stationary distributions which are \emph{proportional} to an objective function, so constant factors can be ignored.  

\begin{figure}[H]
\centering
    \includegraphics[width=0.5\linewidth]{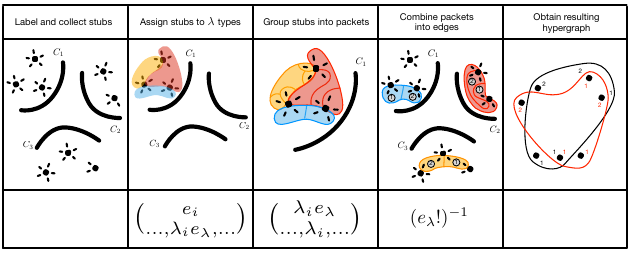}
    \caption{Visualization of each of the constituent counting steps for Equation~\ref{eq.degree.adjusted.entropy}  }
    \label{fig:stubs_and_packets}
\end{figure}

\subsection{Simulated Annealing}

%\pc{revised subsection 2.6}

Our aim is to cluster a hypergraph $\Hy$ by selecting the partition $\braces{C_i}_{i=1}^m$ which maximizes the mutual information between $\Hy$ and the compression $\gamma$ induced by $\braces{C_i}_{i=1}^m$. 
For this section, it is convenient to instead use the vector representation  $\vc \in \Z^n$, where $c_j$ gives the cluster to which node $j$ is assigned by the partition $\braces{C_i}_{i=1}^m$.  
A choice of $\vc$ is equivalent to a choice of partition $\braces{C_i}_{i=1}^m$ and therefore to a choice of compression $\gamma$. 
Hence, we can define the entropy $H(\vc)$ and number of compatible hypergraphs $Z(\vc)$. 
We aim to minimize $Z(\vc)$. 
Performing this minimization exactly is computationally intractable, even for dyadic networks.\cite{rosvallInformationtheoreticFrameworkResolving2007} 
We therefore perform approximate stochastic optimization via simulated annealing.\cite{kirkpatrick1983optimization} 

To perform simulated annealing, we use the Metropolis-Hastings algorithm\cite{chib1995understanding} to construct a random walk on the space of candidate clusterings. 
We begin at a uniformly random clustering $\vc^{(0)} \in \Z^n$.
At each timestep $t$, we select a node and candidate label $(v,i) \in  V\times \braces{1,\dots,m}$ uniformly at random and propose a new state state $\vc'$ where $c_u'=c_u^{(t)}$ for $u \neq v$ and $c_v'=i$. 
Let $\Delta(\vc', \vc) = \log Z(\vc') - \log Z(\vc)$. 
We accept $\vc'$ as the new state with probability $\min\braces{1,e^{-\beta \Delta(\vc', \vc)}}$ and reject $\vc'$ otherwise, where $\beta \geq 0$ is an \emph{inverse temperature} parameter. 
If $\vc'$ is accepted, then we set $\vc^{(t+1)} = \vc'$. 
From standard results on the Metropolis-Hastings algorithm, this random walk has a stationary distribution and the mass of this distribution at $\vc$ is proportional to $Z(\vc)^{-\beta}$. 
The mode(s) of this distribution occur at the value(s) of $\vc$ that minimize $Z(\vc)$, with the sharpness of these modes depending on the inverse temperature $\beta$. 
For small $\beta$, much of the probability mass of the stationary distribution lies away from the modes, whereas as $\beta \rightarrow \infty$ the mass concentrates on these modes. 
In simulated annealing, we allow $\beta = \beta(t)$ to depend on the timestep, gradually increasing $\beta(t)$ as the algorithm proceeds. 

Since we aim to find minima rather than sample from the stationary distribution, we track of the cluster assignment vector that minimizes entropy along our random walk. 
For  pseudocode, see Algorithm \ref{alg:annealing}.

\begin{algorithm}[H]
\caption{This algorithm will use simulated annealing to find a cluster assignment with low entropy. Note that $Z$ implicitly depends on the hypergraph $\Hy$.}\label{alg:annealing}
\begin{algorithmic}[1]
\Procedure{run\_chain}{$\Hy$, number\_of\_clusters, number\_of\_steps, $\beta(t)$}

\State  $\vc\in \braces{1,\dots,\text{number\_of\_clusters}}^{|V(\Hy)|}$ \hfill initialize a random cluster assignment vector
\State best\_entropy $\gets \log Z(\vc)$\hfill initialize the best entropy seen
\State best\_cluster $\gets \vc$ \hfill initialize the best cluster seen
\State $t\gets 0$\hfill initialize the number of steps attempted
\While{ $t < $ number\_of\_steps:}
    \State propose $\vc'$ \hfill randomly choose a neighbor of $\vc$ 
    \State $\Delta \gets \log Z(\vc')-\log Z(\vc)$ \hfill calculate the change in entropy
    \State sample $X\sim U(0,1)$
    \If{$X < \min\braces{1, e^{-\beta(t)\cdot\Delta(\vc', \vc)}}$} \hfill compare $X$ to acceptance probability
    \State $\vc\gets \vc'$ \hfill accept new assignment vector
    \If{ $\log Z(\vc) < $ best\_entropy} \hfill compare new entropy to best\_entropy
    \State best\_entropy $\gets \log Z(\vc)$ \hfill replace best entropy 
    \State best\_cluster $\gets \vc$ \hfill replace best cluster
    \EndIf
    \EndIf
    \State $t\gets t+1$ \hfill increment $t$
\EndWhile\\
\hspace{1em} \Return best\_cluster, best\_entropy \hfill return the best cluster assignment found

\EndProcedure
\end{algorithmic}
\end{algorithm}

\subsection{Model Selection}
The proposed clustering procedure here requires a given number of clusters. While there may be \textit{a priori} well-reasoned choices for sensible values of $m$, the number of clusters to cluster a given hypergraph into, there is no guarantee the interested practitioner will have a selected $m$ in mind. Should $m$ be difficult to choose or unknown \textit{a priori}, we then find ourselves faced with a model selection problem before we may even begin clustering. 

While a variety of approaches have been proposed for choosing the optimal number of communities to cluster a (hyper)graph with have been proposed,\cite{fortunatoCommunityDetectionGraphs2010} an exceptionally simple approach can be found in information theory once again; that is, utilizing the principle of parsimony and choosing an appropriate number of clusters $m$ given the clustering's description length.\cite{hansen2001model, grunwald2007minimum, rosvallInformationtheoreticFrameworkResolving2007}
Similar to the corresponding work by Rosvall and Bergstrom,\cite{rosvallInformationtheoreticFrameworkResolving2007}  we claim a reasonable choice for $m$, unless otherwise constrained by domain knowledge or hypothesis, can be found with the value of $m$ that yields a total description length that is minimal. 
If we express by $L(\Hy)$ the total number of bits to precisely describe $\Hy$, then we can decompose $L(\Hy)$ as
\begin{equation}\label{eq:mdl_general}
    L(\Hy) = L(\gamma) + L(\Hy \mid \gamma)
\end{equation} 
where $\Hy$ is a given hypergraph and $\gamma$ is a proposed compression of $\Hy$.
Hence, our model selection can be performed via the entropy-parsimonious minimum description length value for $m$ given as the solution to the equation
\begin{equation}\label{eq:mdl_optim}
    m^* = \underset{m}{\text{argmin}} \, \Big[ L(\gamma) + L(\Hy \mid \gamma_m) \Big]
\end{equation}
where $\gamma_m$ is the optimal compression of $\Hy$ into $m$-many clusters with our proposed method.

We expand \cref{eq:mdl_general} as

\begin{equation}\label{eq:mdl_expanded}
    L(\gamma) + L(\Hy\mid\gamma) = n\log m + \sum\limits_{k=2}^{k^*}  \binom{m + k - 1}{ k} \log \ell_k +  H(\Hy \mid \gamma)
\end{equation}
where $n$ is the number of nodes in the hypergraph, $m$ is the number of groups in partition $\gamma$, $\ell_k$ is the number of hyperedges of size $k$, and $k^*$ is the size of the largest hyperedge in $\Hy$.

\begin{figure}[H]
    \centering
    \begin{tabular}{c|c}
    \includegraphics[scale = .4]{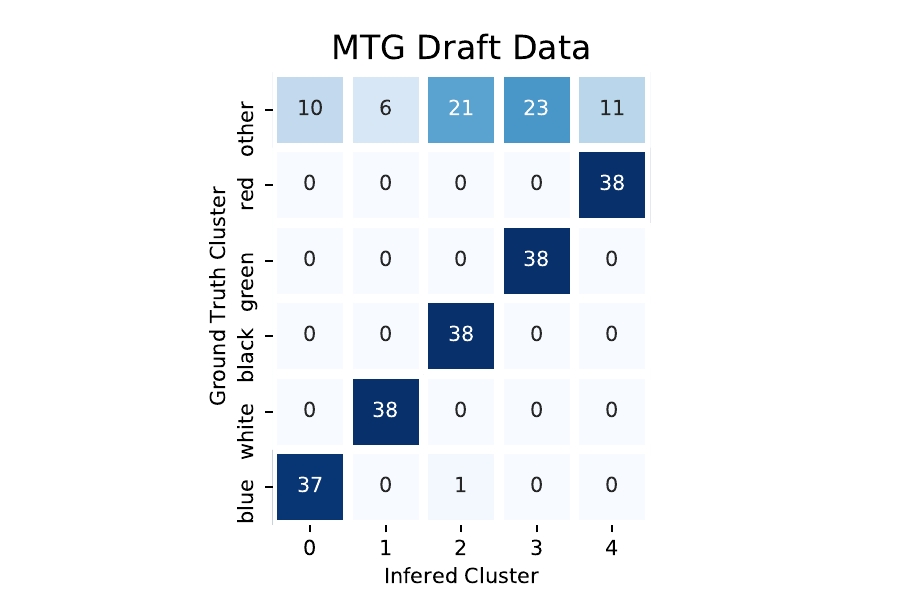}&
    \includegraphics[scale = .4]{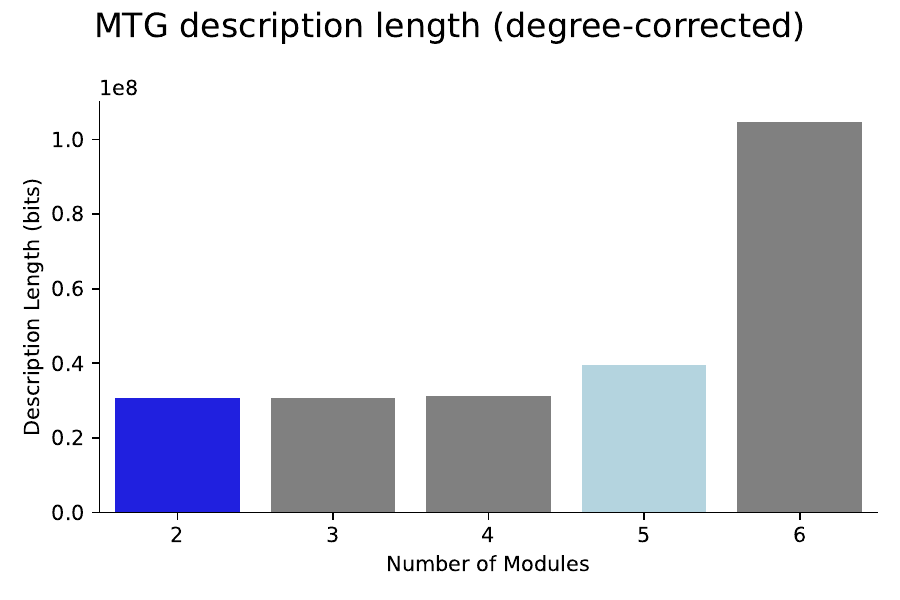}
    %&\includegraphics[scale = .4]{images/MTG_non-degree-adjusted.pdf}  
    \end{tabular}
    \caption{Results from running the degree-corrected chain on the Dominaria United Premier Draft data for $10,000$ steps. 
    The clustering algorithm was looking for 5 clusters, which we break apart into the 5 mono-colored classes and everything else.}
    \label{fig:MTG_5_cluster}
\end{figure}

The description length under this coding scheme is known to be an imperfect procedure for selecting the number of clusters. Indeed, Rosvall and Bergstrom found that it often underestimated the number of clusters as compared to the known value within a given generative model.\cite{rosvallInformationtheoreticFrameworkResolving2007} Likewise, we find that the procedure is not always capable of selecting the known true number of clusters; however, it is capable, nonetheless, of providing some amount of insight and acting as a counterweight to uninformed prior selection to the number of clusters. We show in Figure~\ref{fig:MTG_5_cluster} the description length calculations for the Magic: the Gathering dataset we discuss further in \Cref{sec:empirical}.

\section{Results: Synthetic data} \label{sec:synthetic}

The stochastic block model is a method to generate random graphs with seeded community structure.
Given vertex sets $V_1, \cdots, V_m$ with sizes $n_1,\cdots, n_m$ respectively, we want to generate a hypergraph on the vertex set $\bigcup_{1\leq i \leq m} V_i$, where each $V_i$ is a seeded community within the graph. 
In order to do this, we add a hyperedge of $\lambda$-type with probability $P_\lambda$. 
Communities may be denser or sparser depending on the choice of the probabilities $P_\lambda$.
One motivating idea for the stochastic block model is that every sufficiently large simple graph looks like some stochastic block model with significantly fewer parts than there are vertices (via Szemer\'edi's Regularity Lemma).

We generate hypergraphs according to the following parameters: two ground truth communities of size $n=200$, where each vertex sees on average five $2$-edges and five $3$-edges. 
This means we must generate $5n$ $2$-edges and $\tfrac{10}{3}n$ $3$-edges. 
We generate these edges so that the total proportion of $2$-edges within one of the two seeded clusters is $p_2$ and the total proportion of $3$-edges within one of the two clusters is $p_3$, for various choices of $0\leq p_2,p_3\leq 1$. 
Due to the concentration around the mean, this model roughly corresponds to choosing $P_{(0,2)}=P_{(2,0)}$ with $P_{(0,2)}+P_{(1,1)}=\tfrac{5}{2n}$, and 
$P_{(3,0)}=P_{(0,3)}$, $P_{(2,1)}=P_{(1,2)}$ with $P_{(3,0)}+P_{(1,2)} = \tfrac{10}{n^2}$ where a bit more care needs to be taken to balance the number of edges within communities and between communities. 
The advantage of not strictly following the stochastic block model is that synthetic hypergraphs can be generated in linear (in the number of vertices) time rather than cubic time.

\begin{figure}[H]
    \centering
    \begin{tabular}{c|c}
         \includegraphics[scale= .5]{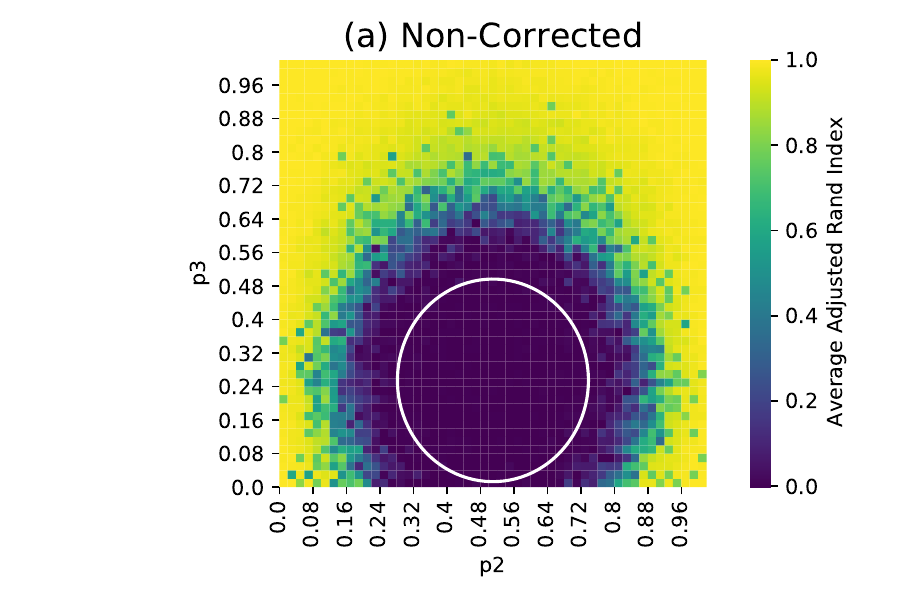}& \includegraphics[scale= .5]{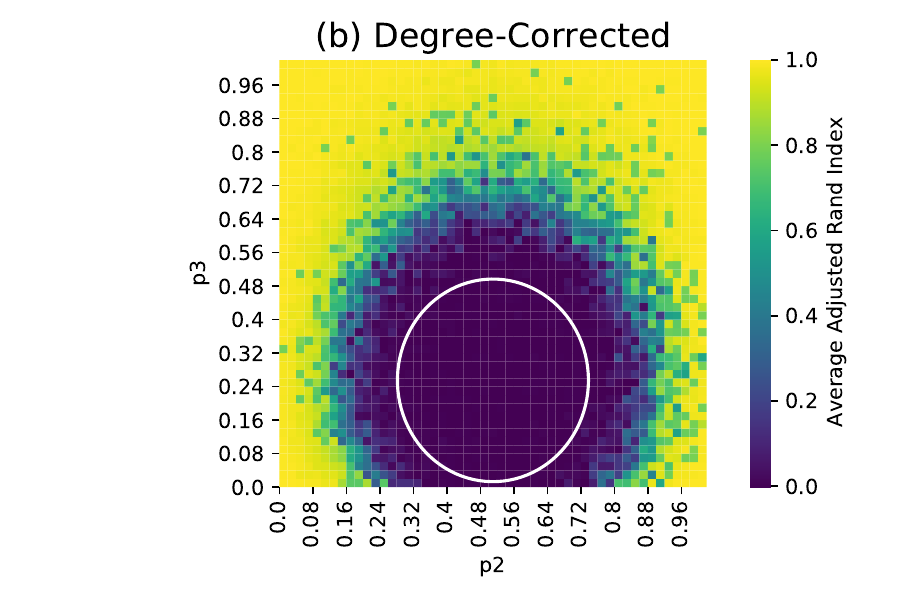} \\
         \hline
         \includegraphics[scale= .5]{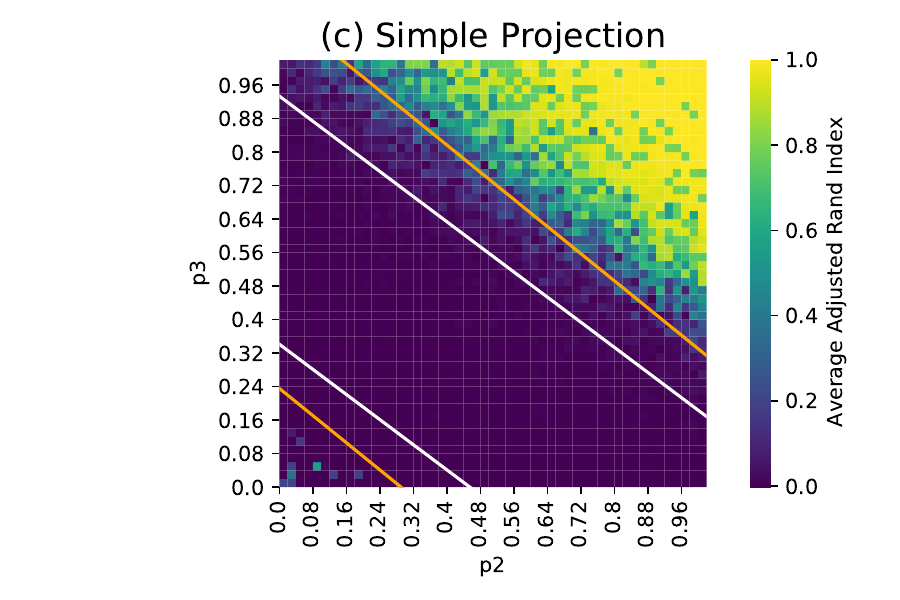}& \includegraphics[scale= .5]{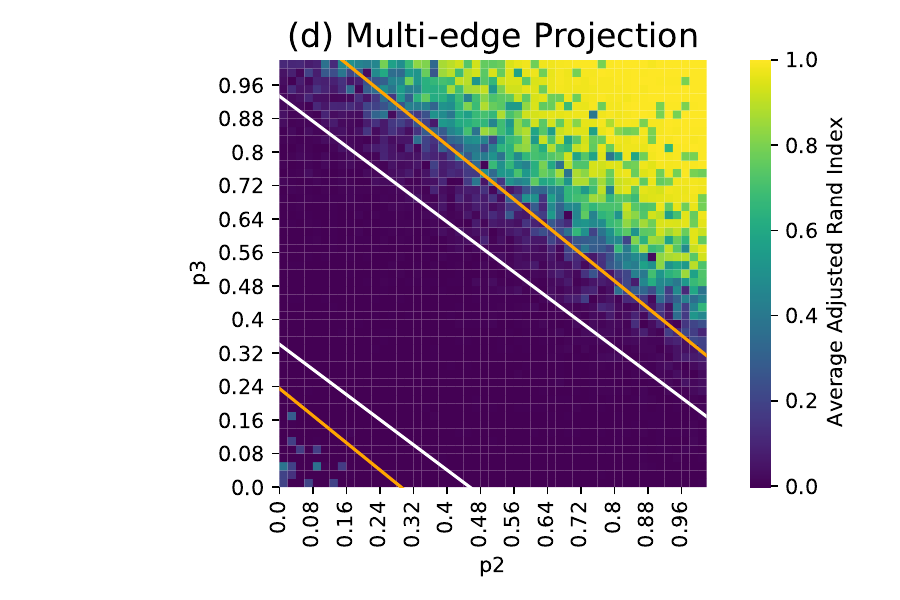}
    \end{tabular}% we can switch to .pdf instead of .pdf for ArXiv
    
    \caption{Each heatmap has $51\times 51$ pixels, where each pixel represents the average ARI across 5 hypergraphs. Each hypergraph underwent 20 independent clustering attempts, of which we used the results from the run which achieved the lowest entropy. The same hypergraphs are used across all four plots.
    The white ellipse in plots (a) and (b) are the conjectured detection threshold for Belief-Propagation Spectral Clustering for hypergraphs. \cite{chodrowNonbacktrackingSpectralClustering2022} 
    The white lines in plots (c) and (d) are the conjectured detection threshold for Non-Backtracking Spectral Clustering for hypergraphs. \cite{chodrowNonbacktrackingSpectralClustering2022} 
    The orange lines in plots (c) and (d) are the proven detection threshold for the graph stochastic block model~\cite{krzakalaSpectralRedemptionClustering2013} for edge densities of the multi-edge projection parameterized by $p_2$ and $p_3$.}
    \label{fig:heatmap}
\end{figure}

The heatmaps in Figure~\ref{fig:heatmap}  show the results of a series of experiments on the planted partition model described above. Each pixel gives the average Adjusted Rand Index (ARI) of the cluster assignments found by our algorithm compared to the planted partition after 20 attempts, for varying parameters of $p_2, p_3$.
In these visualizations, the region bounded by the white curves is the detectability threshold for hypergraph spectral methods conjectured by Chodrow, Eikmeier, and Haddock.\cite{chodrowNonbacktrackingSpectralClustering2022} 
While our present results fall short of these conjectured thresholds, we note that these thresholds were derived under the assumption that certain relevant parameters of a generative model are known exactly. 
In contrast, our proposed method does not require any parameters to be known prior to inference.

\begin{figure}[H]
\centering
    \includegraphics[width=0.5\linewidth]{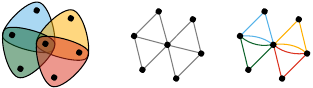}
    \caption{From left to right, we have a hypergraph, its simple clique projection, and its multi-clique projection.}
    \label{fig:expansion}
\end{figure}

We compare the performance of our algorithm on these planted-partition hypergraphs to its performance on the simple and multi-edge projections. 
Recall that the simple projection of a hypergraph is a dyadic graph on the same vertex set, wherein a simple pairwise edge connects each pair of nodes that participate together in some hyperedge.
The projection is a lossy representation of a hypergraph since two vertices are connected by at most one dyadic edge, whether they participate in one hyperedge together or many.
For this reason, we also consider the multi-edge projection, wherein a pair of vertices that participate in $k$ distinct hyperedges are connected by $k$ dyadic edges in the expansion (or equivalently, a single dyadic edge with edge weight $k$).
See Figure~\ref{fig:expansion} for an example.

Plots (c) and (d) in Figure~\ref{fig:heatmap} shows the results of our algorithm using the simple and multi-edge projections, respectively.
The orange lines are the detection thresholds for the graph stochastic block model\cite{krzakalaSpectralRedemptionClustering2013} using the edge densities of the multi-edge projection parameterised by $p_2$ and $p_3$. 
Since the hypergraphs we generated are sparse, there should be relatively few multi-edges in the multi-edge projection, suggesting that the edge densities in the multi-edge and simple projections are similar. 
This also justifies using the detection threshold for the graph stochastic block model, which holds both for sparse and dense simple graphs. 
Interestingly, both the simple and multi-edge projection find some success within the detection threshold, suggesting that some mutual information clustering may be sensitive to some of the latent hypergraph information in the projections. 
For example, the presence of triangles in the projections of a sparse hypergraph are potentially distinguishing from the graph stochastic block model or the sparse Erd\H{o}s-R\'enyi random graph.

\begin{figure}[H]
\centering
    \begin{tabular}{c|c}
    \includegraphics[width=0.45\linewidth]{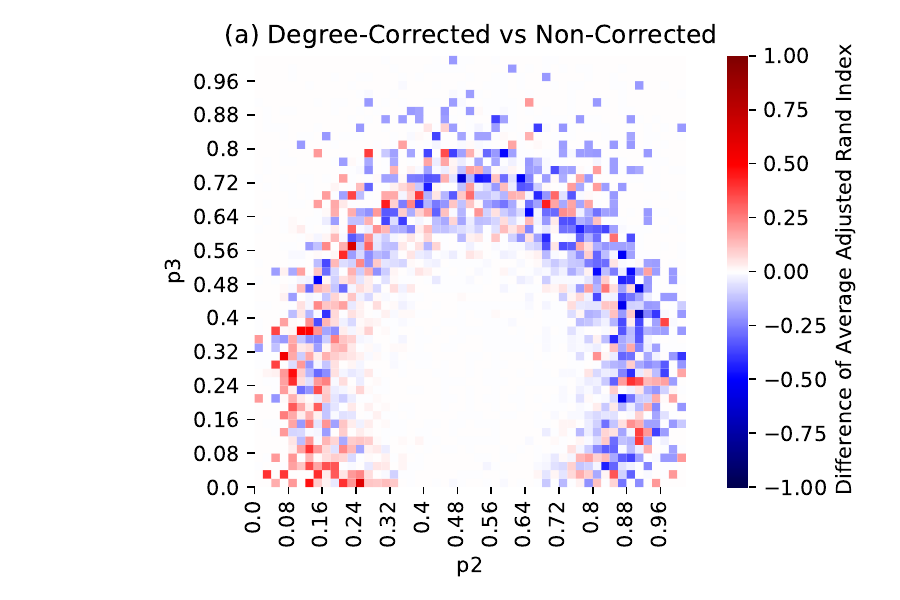}&
    \includegraphics[width=0.45\linewidth]{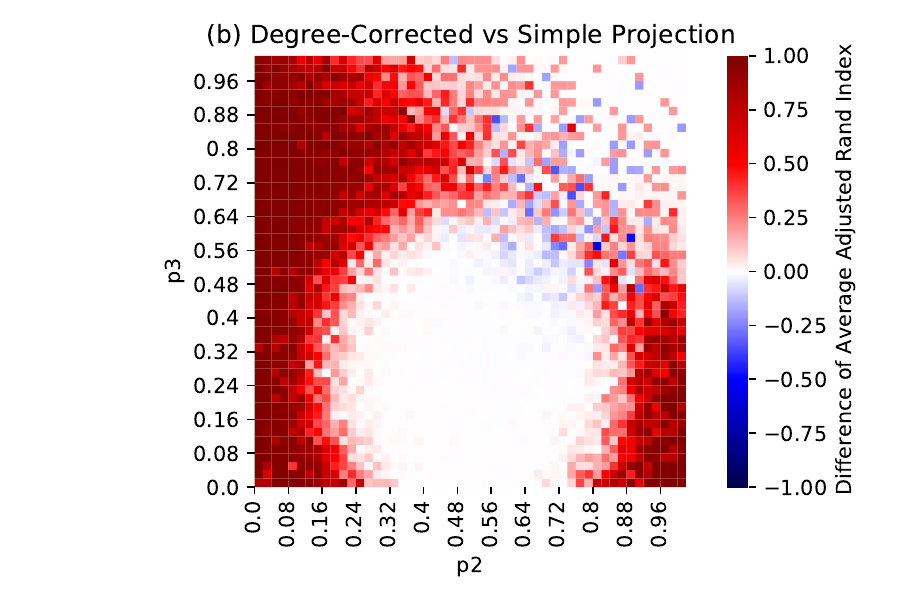}\\
    \hline\includegraphics[width=0.45\linewidth]{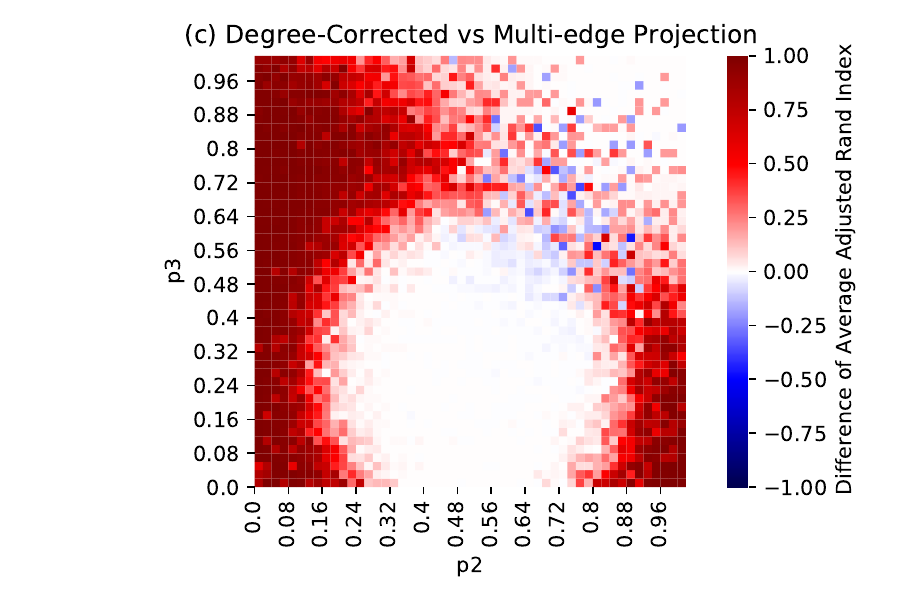}
    \end{tabular}
    \caption{Plot (a) compares the degree adjusted chain against the non-adjusted chain. Interestingly, $p_2,p_3\in [0.00,0.32]$ range appears to favor the degree adjusted chain, while the rest of the circular boundary region favors the non-adjusted chain. 
    Plots (b) and (c) are a comparison of hypergraph native results with multi-edge and simple projection results; red indicates a superior performance by the degree-corrected chain, while blue indicates that the projection performed better.  }
    \label{fig:comparison_heatmap}
\end{figure}

\section{Results: Experiments on Data} \label{sec:empirical}

\subsection{Primary School Contact Hypergraph}

\label{subsec:PSC}

\begin{figure}[H]
    \begin{center}
    \begin{tabular}{c|c}
    \includegraphics[scale = .5]{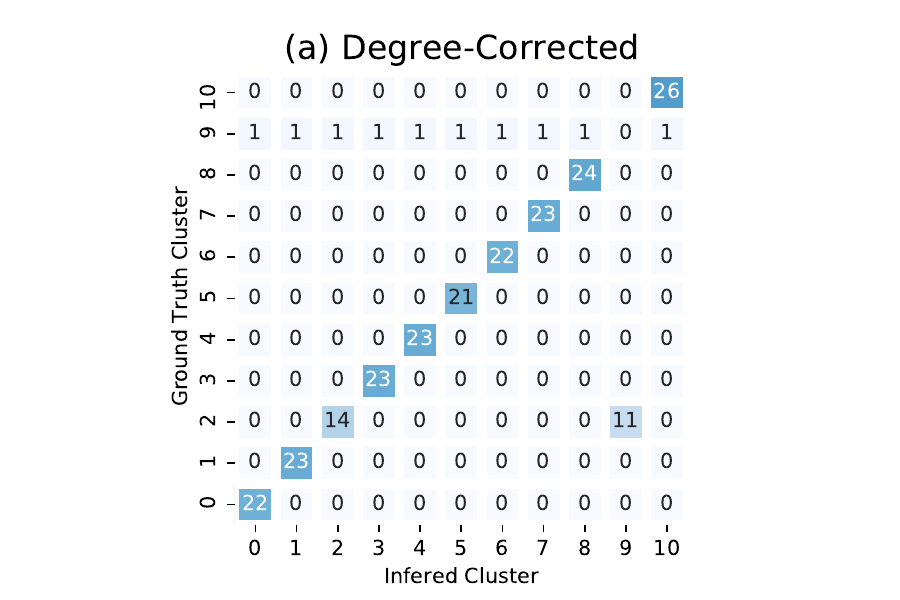}&\includegraphics[scale = .5]{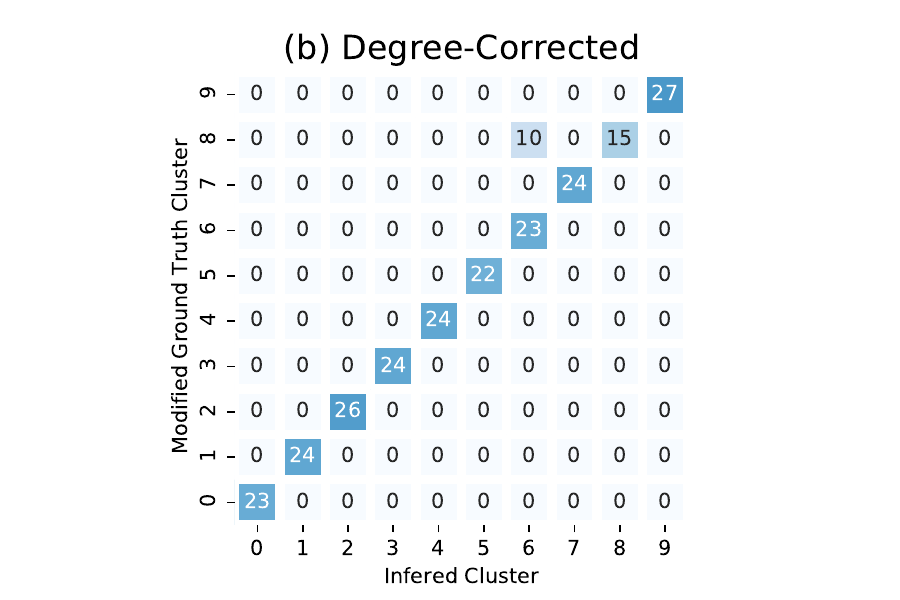}\\\hline
    \includegraphics[scale = .5]{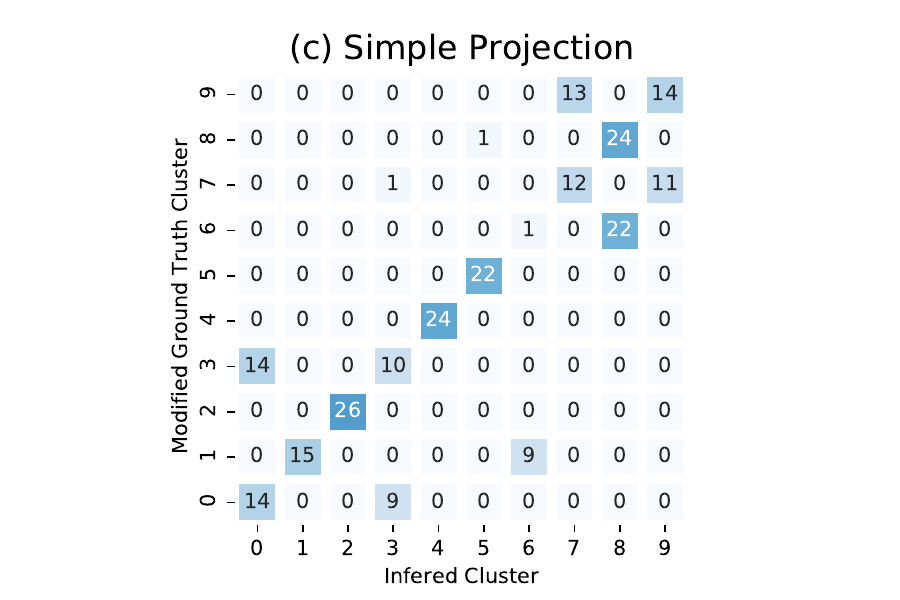}
    &
    \includegraphics[scale = .5]{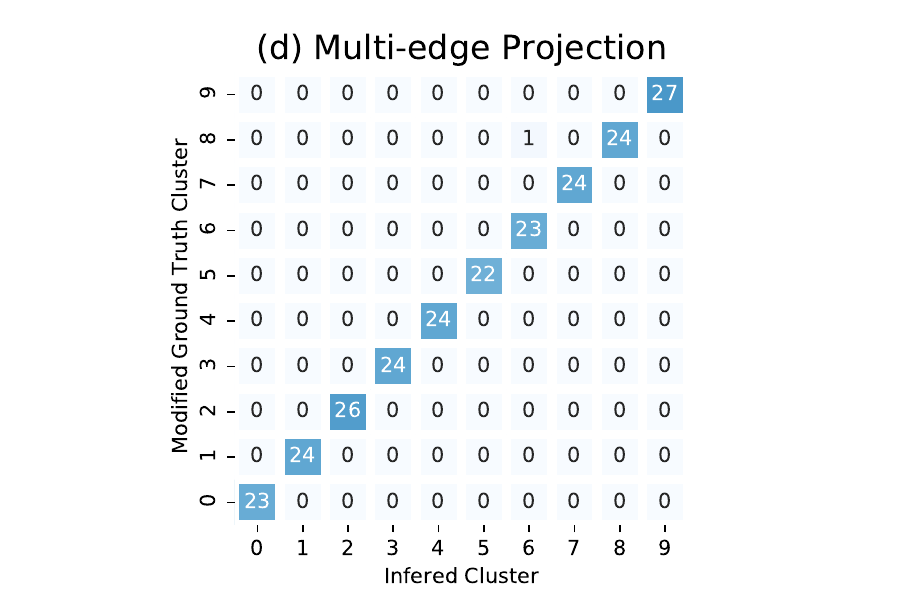}
    \end{tabular}
    \end{center}
    
    \caption{Inferred and ground-truth clusters for the primary school contact data set.\cite{stehleHighResolutionMeasurementsFacetoFace2011} 
    Each matrix is the lowest entropy of 50 independent runs of 20,000 steps.
    The cluster heatmap for (a) uses collected ground truth; this includes a cluster label for teachers within the primary school.
    The cluster heatmaps for (b), (c), and (d) is compared against a modified ground truth where the ``teachers'' cluster from the ground truth is divided up among the classrooms according to the assignments in (a). The ARI values are (a) 0.91, (b) 0.93, (c) 0.64, and (d) 0.99.}
    \label{fig:primary}
\end{figure}

The primary school contact data set obtained from Stehl\'e et al. \cite{stehleHighResolutionMeasurementsFacetoFace2011} provides a hypergraph with 242 vertices. 
Edges in this hypergraph correspond to groups of students and teachers that were within 1.5 meters of each other and facing each other.
The ground truth for this data set assigns students to one of $10$ classrooms, while teachers are all assigned to their own cluster. 
Running our algorithm on this data using 11 clusters resulted in an ARI of $0.91$ after selecting the lowest-entropy cluster assignment from 50 runs with $20,000$ steps each (\Cref{fig:primary}(a)). 
We also studied our model's performance on a modified version of the data set in which each teacher node is given the label of their classroom, resulting in 10 clusters. 
Running our algorithm on the modified data set gave perfect cluster recovery with an ARI of $0.93$, again after $50$ runs with $20,000$ steps (\Cref{fig:primary}(b)) 

We compared our algorithm to two simulated annealing algorithms defined on projections of the data. 
A chain on a simple graph projection obtained an ARI of $0.64$ (\Cref{fig:primary}(c)), while a chain defined on a multi-edge weighted projection misclustered a single element, yielding an ARI of $0.99$ (\Cref{fig:primary}(d)). 
These results indicate the value of higher-order relationships in clustering hypergraph data, and are qualitatively aligned with prior hypergraph algorithms applied to this data set.\cite{chodrow2021generative}

The multi-edge projection out performed the simple projection, and scored nearly as well as the degree-adjsted hypergraph chain. 
This is particularly interesting in the context of our results on synthetic data, where there appears to be no significant difference between simple and multi-edge projections.
One possible explanation is that multi-edge projection of the synthetic hypergraphs produces a simple graph since  sparse hypergraphs have few overlapping hyperedges.

\subsection{High School Contact Hypergraph}

\label{subsec:HSC}

The High School Contact data set produced by Mastrandrea et al.\cite{mastrandrea2015contact} provides a hypergraph with 327 vertices. 
Edges in this hypergraph correspond to groups of students that were within 1.5 meters of each other and facing each other.
The ground truth for this data set assigns students to one of $9$ classrooms. 

\begin{figure}[H]
    \centering
    \begin{tabular}{c|c}
    \includegraphics[scale = .5]{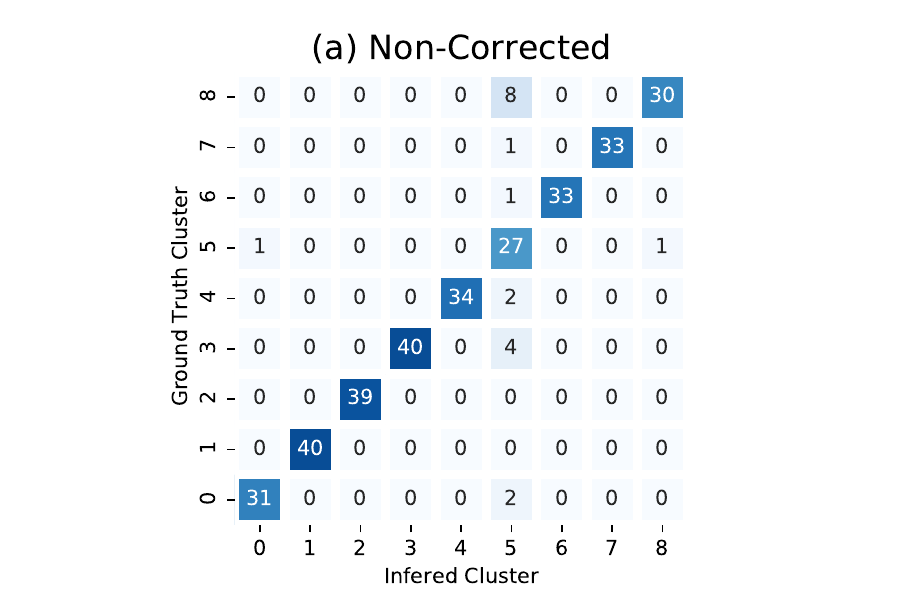} &\includegraphics[scale =.5]{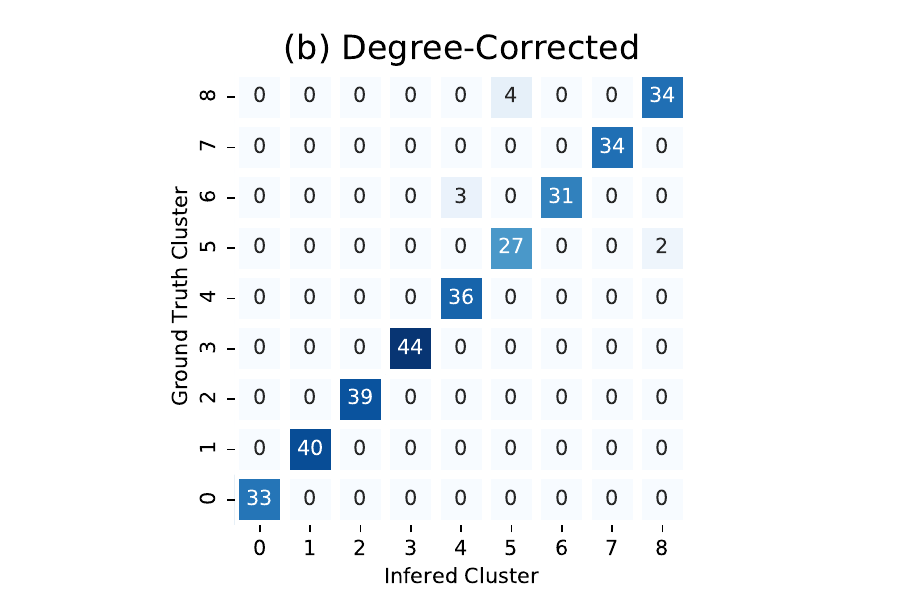} \\\hline\includegraphics[scale = .5]{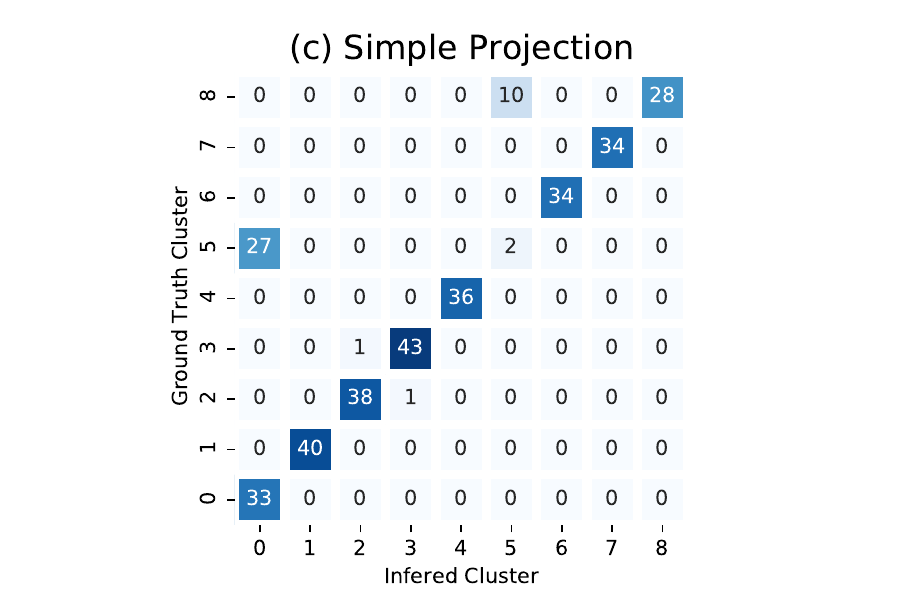}
    & \includegraphics[ scale = .5]{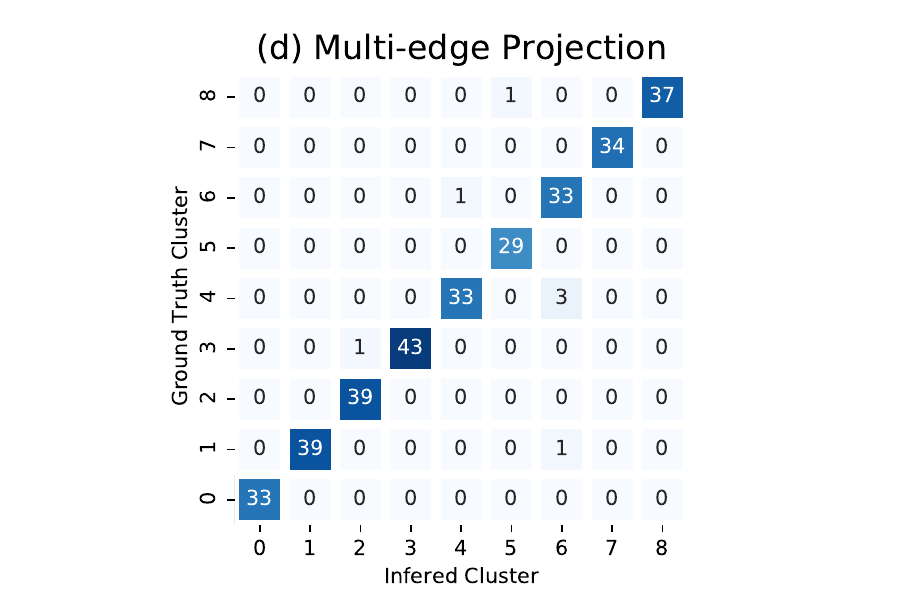}
    \end{tabular}

    \caption{As in \Cref{fig:primary}, using the high-school contact data set.\cite{mastrandrea2015contact} The ARI values are (a) 0.87, (b) 0.95, (c) 0.87, and (d) 0.95.}
    \label{fig:highschool}
\end{figure}

Preliminary exploration of this data set found that the non-adjusted chain did not recover the ground truth clusters. 
For a representative illustration of the performance of the non-corrected chain, see \Cref{fig:highschool}(a). 
In that particular experiment, we obtained an ARI of 0.87 by selecting the lowest entropy observed across 50 independent runs with 20,000 steps each. 

In order to improve our algorithm, we added a degree correction to our objective function. 
These details are presented in Section \ref{sec.deg.adjustement}. 
The degree adjustment leads to better community detection.
This is illustrated by the cluster heat map (b) in Figure~\ref{fig:highschool}, which achieves an ARI of 0.95.

As with the Primary School Contact Data, we ran the degree adjusted chain on the simple and multi-edge projections of the data set. 
The multi-edge  projection performed comparably to the degree adjusted chain with an ARI of $0.95$, and significantly out performed the simple projection which scored an ARI of $0.87$.

The mis-classifications of the simple projection look systematic.
Cluster 5 is combined with cluster 0, while cluster 8 is essentially split into two clusters. 
One possible explanation is that the loss of information about the frequency of interactions between students is lost in the simple projection (as compared to the hypergraph or multi-edge projection).

\begin{figure}[H]
    \centering
    \includegraphics[scale=1]{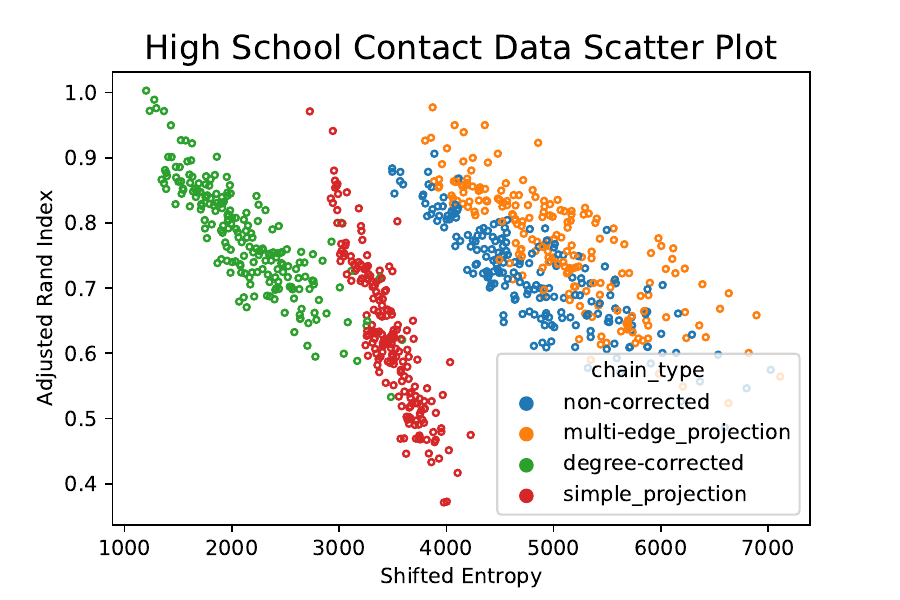}
    \caption{Each scatter plot consists of 200 points and plots the entropy against the ARI. 
    Each point is obtained by running the corresponding chain for 20,000 steps and keeping the lowest entropy observed on that run, then the ARI is calculated using the corresponding clustering.
    Entropy values are shifted to fall within the interval $[0,10000)$.}
    \label{fig:scatter}
\end{figure}

Though the clusterings in Figure~\ref{fig:highschool} suggest that the degree-corrected and the multi-edge projection chains are comparable, there is evidence to suggest that the degree-corrected chain is better. 
The scatter plot in Figure~\ref{fig:scatter} suggests that the degree-corrected chain has the best chance of finding the ground truth clustering, as compared to the other chains. 
Furthermore, the scatter plots show that the entropy of a clustering is inversely correlated with the ARI. 
Notably, out of the 200 attempts with both the degree-adjusted and non-adjusted chains, the highest ARI is achieved by the run with the lowest entropy.

\subsection{Clustering Magic: the Gathering Cards}

\label{subsec:MTG}

Magic: the Gathering draft is a trading card game where eight players open randomized packs of cards and take turns picking cards in a hidden draft. 
After picking 45 cards, players build 23 card decks with which they compete. 
Cards have associated colors; either black, blue, green, red, white,  or any subset thereof (including the empty subset). 
Due to the mechanics of the game, it is typically extremely disadvantageous to have cards from more than 2 color classes in a deck. 
This gives players an incentive to draft their cards concentrated around a pair of colors (for example, one player may concentrate on drafting only white, red, and white-red cards). 

The Magic: the Gathering drafting community collects data on the outcomes of online drafts and the subsequent games. 
This data is publicly available through 17Lands.com. \cite{17Lands}
In particular, we used the Dominaria United Premier Draft data, which contains the card names (including multiplicity) of all the cards in a player's card pool after a draft.
We ignored the multiplicity to make a hypergraph where the vertex set is the set of all cards that could possibly be drafted, and a hyperedge is a player's card pool (without multiplicity) after a draft. 
We ran two experiments with this data. 

In the first experiment, we clustered the hypergraph into 5 clusters assuming that a reasonable ground truth would be the colors of the cards. 
Multi-colored and colorless cards make this notion of ground truth ambiguous. 
Therefore, we scored the clustering only on how the mono-colored cards are partitioned. 
The algorithm only mis-classifies a single mono-colored card; The card ``Coral Colony'' is a blue card that gets clustered with black cards. 

The second experiment applied the minimum description length criterion to determine the number of clusters that are present in the hypergraph. 
This is motivated by the fact that choosing 8 clusters for the clustering algorithm reveals different deck archetypes. 
In particular, there are some multi-color strategies that require certain key cards to enable them. 
Recognizing these archetypes as the ``themes" of the clusters requires some domain knowledge, and is therefore, hard to verify independently. 
However, it does suggest that the minimum description length could reveal a ``better'' ground truth than card color classes. 
Unfortunately, our experiment testing different cluster numbers suggests that the 2 clusters provide the shortest description length.

\section{Discussion} \label{sec:discussion}

In this document, we establish a novel information-theoretic framework for clustering hypergraph data generalizing the graph theoretic framework established by Rosvall and Bergstrom~\cite{rosvallInformationtheoreticFrameworkResolving2007} while incorporating degree correction methods driven by stochastic blockmodel generative models in the style of Piexoto.~\cite{peixotoNonparametricBayesianInference2017} 
We have found that our algorithm is able to recover structures in synthetic and real-world hypergraphs, with performance that is often competitive with methods based on projections of dyadic graphs. 
Importantly, we find that degree correction leads to substantial improvements over non-degree-corrected methods on empirical data sets. 
We also offer a method based on minimum description-length (MDL) for estimating the appropriate number of communities in data when this is not known \emph{a priori}. 

Our results pose several directions of future work. 
First, our algorithm for clustering is relatively slow. 
This is due in part to the complicated, highly nonconvex structure of the energy landscape of the entropy minimization objective. 
Furthermore, our proposed algorithm considers only single-node transitions between cluster labels. 
Merge-split methods such as those discussed by Peixoto (2020) for dyadic graphs may improve performance dramatically.\cite{peixotoMergesplitMarkovChain2020} 
Second, it would be of considerable interest to benchmark our proposed algorithm in both speed and clustering performance against the many existing hypergraph clustering and partitioning methods in a variety of application areas. 
Of special interest are algorithms designed for specific domains, such as balanced partitioning, \cite{schlagHighQualityHypergraphPartitioning2022} image segmentation,\cite{ducournauReductiveApproachHypergraph2012} or circuit design. \cite{karypisMultilevelHypergraphPartitioning1999}
Finally, the framework of data analysis as a compression-motivated optimization problem is one which may have use in other directions. 
Formulating more analysis problems in terms of compression would allow us to deploy combinatorial optimization techniques in the service of complex systems science.

\section{Data Availability}
The data used in Section~\ref{sec:empirical} consists of the Primary School Contact data and  High School Contact data~\cite{mastrandrea2015contact} as well as the Magic: The Gathering data~\cite{17Lands}. The first two are available through Austin Benson's data web page~\cite{BensonPSC},~\cite{Benson}~\cite{BensonHSC}, while the Magic: The Gathering data is directly available through \url{ https://www.17lands.com/}~\cite{17Lands}

\bibliography{one_bib}

\section*{Acknowledgements}
This work was initiated at the 2022 American Mathematical Society (AMS) Mathematics Research Communities (MRC) workshop on \textit{Models and Methods for Sparse (Hyper)Network Science}. We would like to thank the AMS for the opportunity to bring together early-career mathematicians to work on problems related to Business, Industry, and Government. This material is based upon work supported by the National Science Foundation under Grant Number DMS 1916439. Pacific Northwest National Laboratory is operated by Battelle for the DOE under Contract DE-AC05-76RL0 1830. PNNL Information Release PNNL-SA-188428. We are grateful to  Jamie Haddock for useful conversations during the early stages of this work.

\section*{Author Contributions Statement}

All authors participated in the initial AMS MRC working group.
BK proposed and coordinated the project. 
JK led the implementation of algorithms and experiments, with significant contributions from OAR, IA, DK.
FL made Figure~\ref{fig:stubs_and_packets}.
JB, PC, DK, BK, and JK produced the manuscript.
All authors reviewed and edited the final manuscript. 
OAR, IA, JB, TG, FL, and SM contributed equally to this work. 

%JK implemented algorithms, designed experiments, performed experiments, and helped write the manuscript.
%OAR implemented algorithms, performed experiments, and helped write the manuscript. 
%IA implemented algorithms, performed experiments, and helped write the manuscript. 
%DK implemented algorithms, designed experiments, performed experiments, and helped write the manuscript. 
%PC helped to design algorithms, design experiments, and write the manuscript. 
%BK proposed the project, helped to design algorithms, helped to design experiments, and helped to write the manuscript.  

%design: JK, OAR, JB, PC, BK
%implementation: JK, OAR, DK, IA
%experiments: JK, OAR, IA, DK 
%writing: JK, JB, DK, BK, PC, IA
\end{document}